\documentclass{article}

\usepackage{arxiv}

\usepackage[utf8]{inputenc} 
\usepackage[T1]{fontenc}    
\usepackage{hyperref}       
\usepackage{url}            
\usepackage{booktabs}       
\usepackage{amsfonts}       
\usepackage{nicefrac}       
\usepackage{microtype} 
\usepackage{lipsum}
\usepackage[most]{tcolorbox}
\usepackage{graphicx}
\usepackage{multirow}
\graphicspath{ {./images/} }

\title{Progressive Code Integration for Abstractive Bug Report Summarization}

\author{
    Shaira Sadia Karim\textsuperscript{*\dag} \\
    \texttt{shairasadia@iut-dhaka.edu}
    \And
    Abrar Mahmud Rahim\textsuperscript{*\dag} \\
    \texttt{abrarmahmud@iut-dhaka.edu}
    \And
    Lamia Alam\textsuperscript{*\dag} \\
    \texttt{lamiaalam@iut-dhaka.edu}
    \And
    Ishmam Tashdeed\textsuperscript{\dag} \\
    \texttt{ishmamtashdeed@iut-dhaka.edu}
    \And
    Lutfun Nahar Lota\textsuperscript{\dag} \\
    \texttt{lota@iut-dhaka.edu}
    \And
    Dr.\ Md.\ Abu Raihan M.\ Kamal\textsuperscript{\dag} \\
    \texttt{raihan.kamal@iut-dhaka.edu}
    \And
    Dr.\ Md.\ Azam Hossain\textsuperscript{\dag} \\
    \texttt{azam@iut-dhaka.edu}
    \vspace{1.0em} \\
    \textsuperscript{\dag}Islamic University of Technology, Gazipur, Bangladesh \\
    \textsuperscript{*}\,Equal Contribution
}

\begin{document}
\maketitle
\begin{abstract}
Bug reports are often unstructured and verbose, making it challenging for developers to efficiently comprehend software issues. Existing summarization approaches typically rely on surface-level textual cues, resulting in incomplete or redundant summaries, and they frequently ignore associated code snippets, which are essential for accurate defect diagnosis. To address these limitations, we propose a progressive code-integration framework for LLM-based abstractive bug report summarization. Our approach incrementally incorporates long code snippets alongside textual content, overcoming standard LLM context window constraints and producing semantically rich summaries. Evaluated on four benchmark datasets using eight LLMs, our pipeline outperforms extractive baselines by 7.5\%--58.2\% and achieves performance comparable to state-of-the-art abstractive methods, highlighting the benefits of jointly leveraging textual and code information for enhanced bug comprehension.
\end{abstract}


\section{Introduction}
Bug reports are a critical component of software maintenance, providing developers with information about defects, failure conditions, and potential fixes. However, their unstructured nature, redundancy, and occasional ambiguity make them challenging to analyze efficiently \cite{8466000,10.1145/1453101.1453146}. Existing summarization methods, ranging from supervised learning approaches to classical NLP-based techniques, predominantly focus on textual content and often overlook the semantic connections between bug descriptions and their corresponding source code \cite{10.1145/3196321.3196326}. While extractive approaches are widely used, they can produce redundant, fragmented, or incoherent summaries. Abstractive methods, which have the potential to generate more concise and semantically rich summaries, remain underexplored due to their technical complexity \cite{ELKASSAS2021113679,kryscinski-etal-2020-evaluating}.

The advent of Large Language Models (LLMs) has enabled more advanced summarization capabilities \cite{zhang2024systematicsurveytextsummarization,achiam2023gpt}. However, prior LLM-based approaches for bug-report summarization largely ignore the inclusion of full code snippets \cite{7884638}, which are crucial for understanding the context of software defects. LLMs also face challenges related to missing background knowledge, domain-specific terminology, and structured code elements \cite{10.1145/1718918.1718973,FSE086}, although prompt engineering has been shown to mitigate some of these limitations and improve generation quality \cite{10.5555/3495724.3495883,jiang-etal-2020-know}.

To address these gaps, we propose an LLM-based abstractive bug-report summarization framework that integrates textual bug reports with complete code snippets through carefully designed prompt engineering. Our method is evaluated across multiple benchmark datasets, including SDS \cite{10.1145/1806799.1806872}, ADS \cite{ADS}, Fang et al.’s corpus \cite{10.1109/ICSE48619.2023.00060}, and Defects4J \cite{defects4j}. The primary contributions of this work are threefold: (1) the first explicit use of full code snippets to enhance abstractive bug-report summarization, (2) a prompt-engineered framework for effectively combining text and code, and (3) demonstrably improved semantic summary quality as measured by BERTScore \cite{zhang2020bertscoreevaluatingtextgeneration}.

\section{Related Work}
\subsection{Abstractive Bug Report Summarization}
Abstractive summarization of bug reports presents significant challenges due to the technical complexity, domain-specific terminology, and heterogeneous structure of report content \cite{10.1007/s10462-016-9475-9, Shakil_2024}. Traditional methods often struggle to capture the semantic essence of both textual descriptions and associated code, leading to summaries that may omit critical fault-relevant information. 

Recent advances in Large Language Models (LLMs) have substantially improved context-aware summarization capabilities \cite{zhang2024systematicsurveytextsummarization}. Approaches such as RTA \cite{10.1109/ICSE48619.2023.00060}, KSCLP \cite{ksclp}, and SumLLaMA \cite{xiang2024sumllama} leverage pre-training on large corpora, knowledge-specific or contrastive learning techniques, and parameter-efficient fine-tuning to enhance performance on bug-report summarization tasks. While these methods achieve improved fluency and abstraction compared to extractive techniques, they generally focus on the textual content of bug reports and do not incorporate the underlying buggy code \cite{summarizingSoftwareartifacts}. This omission limits the ability of these models to fully capture the context of software defects and constrains their applicability in code-aware summarization scenarios.

\subsection{Prompt Engineering}
Prompt engineering has emerged as a powerful technique for guiding LLMs in complex software engineering tasks \cite{santana2025promptingtechniqueiuse}. Frameworks such as LIBRO \cite{kang2023large}, ChatBR \cite{10764843}, and LLIFT \cite{10.1145/3649828} employ structured prompts, few-shot examples, and progressive prompting strategies to enhance performance on tasks including test-case generation, bug detection, and bug-report generation. These methods leverage the inherent generative capabilities of LLMs while providing explicit cues to shape output content and maintain task relevance.

Similarly, ChatGPT and other general-purpose LLMs have demonstrated effectiveness in vulnerability management and abstractive bug-report summarization \cite{10.5555/3698900.3698946}. However, current prompt-based approaches predominantly focus on textual bug reports and do not incorporate the underlying code associated with defects. This limitation constrains their ability to capture the full semantic and structural context of software bugs, highlighting a gap that motivates the development of code-aware abstractive summarization techniques.

\section{Approach}

\subsection{Problem Statement}
Let $\mathcal{X}_T$ denote the space of textual bug–report utterances (e.g., natural-language descriptions, reproduction steps, error messages) and let $\mathcal{X}_C$ denote the space of code snippets (e.g., stack traces, diffs, method bodies). A bug report is represented as a pair
\[
x = (x_T, x_C) \in \mathcal{X}_T \times \mathcal{X}_C,
\]
where $x_T$ is a sequence of natural-language tokens and $x_C$ is a sequence of code tokens. Formally,
\[
x_T = (t_1, t_2, \dots, t_{n_T}), \qquad t_i \in V_T,
\]
\[
x_C = (c_1, c_2, \dots, c_{n_C}), \qquad c_j \in V_C,
\]
with $V_T$ and $V_C$ denoting the vocabularies for text and code respectively. A summary is an abstractive natural-language sequence
\[
y = (y_1, y_2, \dots, y_m) \in \mathcal{Y},
\]
where $y_k \in V_T$ and $m \ll n_T + n_C$.

We define the \emph{abstractive bug-report summarization problem} as learning a conditional probability distribution
\[
p_\theta(y \mid x_T, x_C),
\]
parameterized by $\theta$, such that generated sequences $y$ satisfy two criteria:

\begin{itemize}
    \item \textbf{Semantic Sufficiency:} The summary $y$ preserves the core meaning of the bug report, capturing essential fault symptoms, reproduction conditions, and implicated code regions. More formally, we assume a semantic representation function
    \[
    \phi : \mathcal{X}_T \times \mathcal{X}_C \rightarrow \mathcal{Z},
    \]
    mapping inputs to a latent semantic space $\mathcal{Z}$, and a corresponding summary-induced representation
    \[
    \psi : \mathcal{Y} \rightarrow \mathcal{Z}.
    \]
    A summary is semantically sufficient if
    \[
    d(\phi(x_T, x_C), \psi(y)) \le \varepsilon,
    \]
    for a task-appropriate similarity metric $d : \mathcal{Z} \times \mathcal{Z} \to \mathbb{R}_{\ge 0}$ (e.g., cosine distance in an embedding space) and a distortion threshold $\varepsilon > 0$.
    
    \item \textbf{Abstractive Novelty:} The summary may contain tokens not present in the original bug report or code snippet. Formally,
    \[
    y \not\subseteq x_T \cup x_C \quad \text{is allowed},
    \]
    distinguishing the task from extractive summarization, which restricts summaries to subsequences or selections from the input.
\end{itemize}

Thus, abstractive summarization requires learning a generative transformation
\[
f_\theta : \mathcal{X}_T \times \mathcal{X}_C \rightarrow \mathcal{Y},
\]
capable of synthesizing novel linguistic expressions while preserving semantic fidelity.

To operationalize this transformation with a Large Language Model (LLM), we construct a structured prompt
\[
\pi(x) = g(x_T, x_C),
\]
where $g$ is a deterministic formatting function (e.g., using delimiters, metadata fields, or role annotations) that exposes salient textual and code structure to the model. The LLM produces a summary
\[
\hat{y} = \arg\max_{y \in \mathcal{Y}} p_\theta(y \mid \pi(x)),
\]
or a sampled approximation thereof.

The objective of this work is to design the prompt representation $g$, specify an effective prompting or training strategy for the LLM, and evaluate the resulting summaries according to both linguistic quality and semantic fidelity with respect to the original bug report and its associated code.

\begin{figure}[htb]
    \centering
    \includegraphics[width=0.90\textwidth]{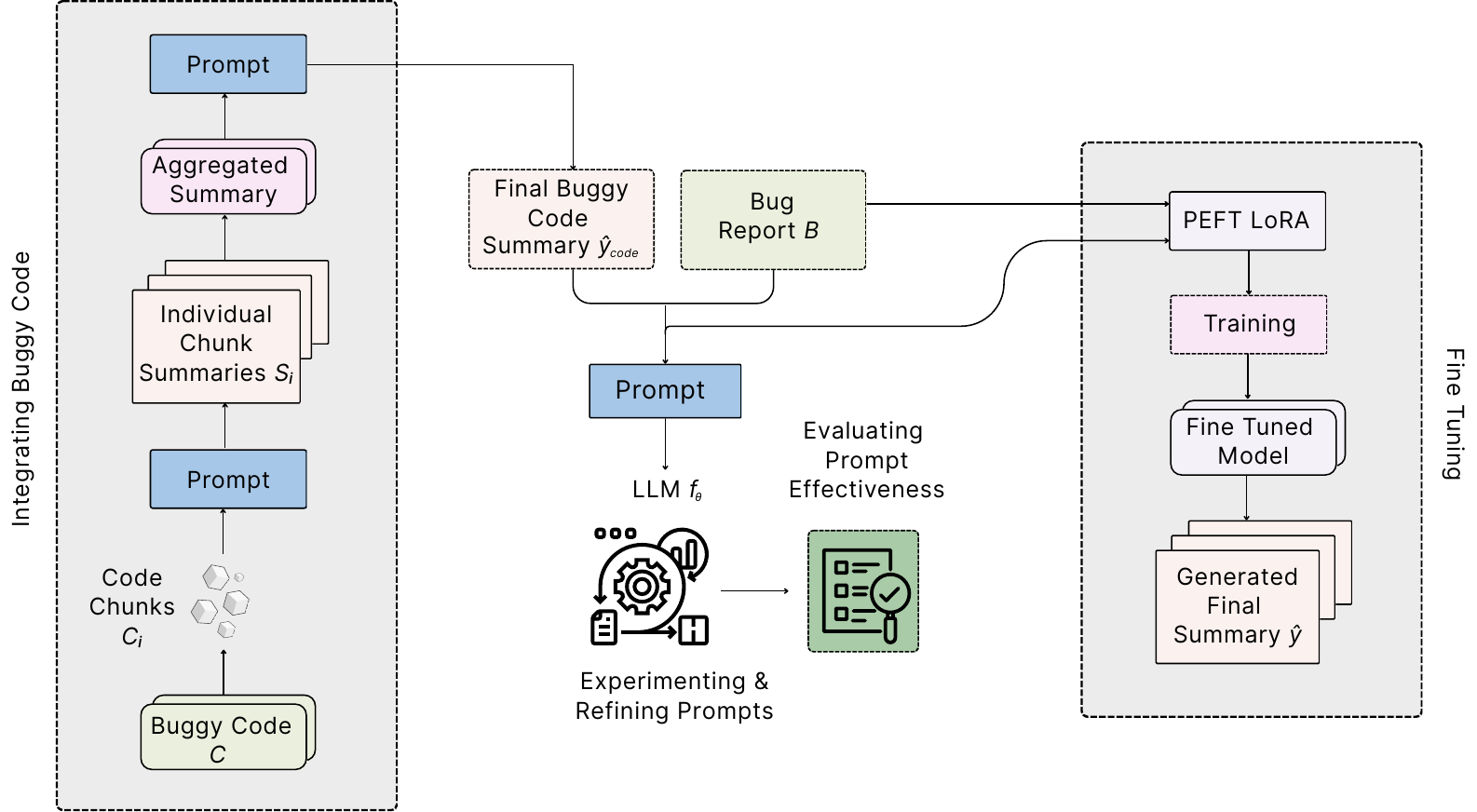}
    \caption{End-to-end pipeline for abstractive bug report summarization. Here long buggy code is chunked and summarized, combined with bug-report text through prompting and refined using LoRA-based fine-tuning to produce summaries.
}
\end{figure}

\subsection{Data Preprocessing}
We construct our dataset by leveraging Defects4J \cite{defects4j}, a curated benchmark suite containing real-world software defects from Java projects, and by augmenting it with bug-report information obtained via a custom web-scraping and data-normalization pipeline \cite{defects4J-dissection}. Formally, let $\mathcal{B}$ denote the set of bug identifiers provided by Defects4J. For each bug $b \in \mathcal{B}$, Defects4J supplies (i) the buggy and fixed program versions and (ii) metadata linking each defect to an external issue tracker.

Our preprocessing pipeline defines a mapping
\[
h : \mathcal{B} \rightarrow \mathcal{X}_T \times \mathcal{X}_C,
\]
which retrieves and normalizes the relevant textual and code modalities for each bug. Concretely, for a given bug $b$:

\begin{itemize}
    \item We scrape and extract the associated natural-language bug report from its corresponding issue tracker, yielding an unstructured text document $r_b$.
    \item We identify code snippets $s_b$ related to the defect by mining commit diffs, stack traces, and referenced file paths from the Defects4J metadata. These snippets often exhibit heterogeneous formats, including inline code blocks, patch fragments, and trace logs.
    \item We apply a normalization operator
    \[
    \nu : \mathcal{R} \times \mathcal{S} \rightarrow \mathcal{X}_T \times \mathcal{X}_C,
    \]
    where $\mathcal{R}$ and $\mathcal{S}$ denote raw textual and code sources respectively, to produce structured sequences of tokens suitable for downstream LLM consumption. This step includes HTML cleaning, tokenization, code-block extraction, Unicode normalization, and removal of incidental noise (e.g., markup artifacts or irrelevant metadata).
\end{itemize}

Thus, each bug $b$ is ultimately mapped to a pair
\[
(x_T^{(b)}, x_C^{(b)}) = h(b),
\]
where $x_T^{(b)} \in \mathcal{X}_T$ is a cleaned and tokenized bug-report text sequence, and $x_C^{(b)} \in \mathcal{X}_C$ is a set of structurally standardized code snippets. This preprocessing pipeline transforms heterogeneous, unstructured data into a consistently formatted representation that is directly usable by LLM-based summarization models.

\subsection{Prompt Engineering}
Our approach employs multi-modal structured prompts that explicitly integrate both textual bug-report content and associated code snippets in order to guide the LLM toward producing coherent and semantically faithful abstractive summaries. Let $\pi(x)$ denote the prompt constructed from an input pair $x = (x_T, x_C)$, where $\pi$ formats the content using templated delimiters, role specifications, and modality markers. This structured formulation provides the LLM with clear boundaries between natural-language descriptions and code-related information, reducing hallucination and improving semantic grounding.

We explored several prompting regimes, including zero-shot, one-shot, and few-shot strategies. In the zero-shot setting, the model receives only the structured prompt and is expected to infer the desired summarization behavior. One-shot and few-shot variants extend the prompt with one or more demonstration pairs $(x^{(i)}, y^{(i)})$, enabling the model to condition on exemplar input–output relationships. These exemplars were selected to minimize domain-specific bias and to expose the model to representative patterns of bug-report language and code structures. Formally, a few-shot prompt can be viewed as the concatenation
\[
\pi_{\text{fs}}(x) 
= \big\langle \pi(x^{(1)}), y^{(1)}, \ldots, \pi(x^{(k)}), y^{(k)}, \pi(x) \big\rangle,
\]
where $k$ determines the number of demonstrations provided.

Long code snippets, which frequently exceed the context-length constraints of contemporary LLMs, required an additional hierarchical processing step. Given a code sequence $x_C = (c_1, \dots, c_{n_C})$ and a maximum permissible context length $L$, we partition the code into contiguous segments
\[
x_C^{(1)}, x_C^{(2)}, \dots, x_C^{(M)},
\quad M = \left\lceil \frac{n_C}{L} \right\rceil,
\]
each satisfying the length limit. For each segment $x_C^{(m)}$, we generate an intermediate summary $s^{(m)}$ by conditioning the LLM on a prompt $\pi\!\left(x_T, x_C^{(m)}\right)$. These intermediate representations are subsequently aggregated into a consolidated code-level summary $\tilde{s}$ through a second-stage abstraction pass, effectively compressing multi-segment code information into a unified semantic description. The final prompt used for generating the complete abstractive summary is then formed by combining the bug-report text $x_T$ with the aggregated code summary $\tilde{s}$, enabling the LLM to operate on a contextually rich yet size-constrained representation.

This hierarchical prompting strategy ensures that both textual and code modalities contribute meaningfully to the final abstractive summary, while also preserving compatibility with LLM context limitations and maintaining high coherence across different stages of generation.

\subsection{Fine-Tuning}
To adapt pre-trained LLMs to the domain of software bug-report summarization, we conducted parameter-efficient fine-tuning using supervised examples derived from the processed dataset. Let each training instance consist of an input pair $x = (x_T, x_C)$ and a corresponding target summary $y$. We constructed two dataset variants to evaluate the role of code information in the learning process. The first variant conditions the model solely on the natural-language bug-report text $x_T$, while the second augments this input with the aggregated code summary $\tilde{s}$, producing an enriched representation $(x_T, \tilde{s})$. This design enables a controlled examination of whether incorporating code-derived semantic content materially improves model performance.

Fine-tuning was performed using Low-Rank Adaptation (LoRA) \cite{hu2021loralowrankadaptationlarge}, a method that inserts trainable low-rank matrices into selected weight components of the transformer architecture. Formally, a weight matrix $W_0 \in \mathbb{R}^{d \times d}$ is reparameterized during fine-tuning as
\[
W = W_0 + BA,
\]
where $A \in \mathbb{R}^{r \times d}$ and $B \in \mathbb{R}^{d \times r}$ are the learned low-rank factors with $r \ll d$. This formulation substantially reduces the number of trainable parameters, allowing efficient specialization of the base model without modifying the full parameter set. LoRA thus offers a computationally tractable way to fine-tune large-scale models while preserving their general linguistic and reasoning capabilities.

Two fine-tuned model variants were produced: one trained on bug-report-only inputs and one trained on inputs enriched with code summaries. These variants allow a direct comparison of the conditional distributions
\[
p_{\theta_{\text{BR}}}(y \mid x_T) \quad \text{and} \quad p_{\theta_{\text{BR+CS}}}(y \mid x_T, \tilde{s}),
\]
thereby quantifying the contribution of code-level context to the abstractive summarization task. The resulting models provide empirical insight into how multimodal conditioning influences the fidelity, conciseness, and technical accuracy of generated summaries.

\subsection{Research Questions}
This work is guided by three central research questions that structure our investigation into LLM-based abstractive bug-report summarization. The first question \textbf{(RQ1)} concerns the comparative effectiveness of our proposed methodology relative to state-of-the-art approaches. We aim to determine whether structured prompting, multimodal integration, and parameter-efficient fine-tuning yield measurable improvements in summary quality when benchmarked against existing extractive and abstractive baselines.

The second question \textbf{(RQ2)} examines the role of code information in shaping summary fidelity and technical accuracy. By contrasting models trained or prompted with bug-report text alone against those additionally conditioned on full or hierarchically summarized code snippets, we seek to evaluate how code-level context influences semantic completeness, specificity, and alignment with underlying defect characteristics.

The third question \textbf{(RQ3)} focuses on the effect of methodological choices—specifically prompt engineering strategies and fine-tuning configurations—on summarization performance across different LLMs and dataset variants. This question addresses the broader issue of how model behavior varies under different adaptation regimes and aims to identify principled strategies for constructing, prompting, and customizing LLMs for software-maintenance tasks.

\subsection{Experimental Setup}

\subsubsection{Dataset}
To evaluate the effectiveness and generalizability of our approach, we conducted experiments on four benchmark datasets that vary in report length, authorship patterns, and overall scale. The Defects4J dataset \cite{defects4j} serves as our primary benchmark for code-aware summarization, providing paired bug reports and corresponding code snippets, which allows for multimodal modeling and hierarchical summarization strategies. The SDS \cite{10.1145/1806799.1806872} and ADS \cite{ADS} datasets are employed as extractive summarization baselines, offering manually annotated summaries or reference selections that enable a comparative assessment of abstractive methods against traditional extractive approaches. Finally, we include the dataset curated by Fang et al. \cite{10.1109/ICSE48619.2023.00060} to test generalization across different projects and reporting conventions. This dataset provides sampled bug-report subsets with high-quality human annotations, facilitating evaluation of model performance in scenarios that differ from those in the Defects4J, SDS, and ADS benchmarks. Collectively, these datasets allow a comprehensive assessment of our models in both multimodal and text-only summarization contexts.

\subsubsection{Models}
We evaluated eight state-of-the-art LLMs spanning a range of sizes, architectures, and levels of code specialization to assess their effectiveness in abstractive bug-report summarization. The selected models include CodeLlama \cite{rozière2024codellamaopenfoundation}, which is explicitly trained for code understanding and generation; Llama-3.1 \cite{grattafiori2024llama} and Mistral \cite{jiang2023mistral7b}, representing large general-purpose LLMs; Phi-3 \cite{abdin2024phi} and Gemma \cite{team2024gemma}, designed for mixed-language reasoning and knowledge-intensive tasks; Qwen3 \cite{yang2025qwen3}, optimized for multilingual and technical text processing; DeepSeek Coder \cite{guo2024deepseekcoderlargelanguagemodel}, focused on code-related tasks; and GPT-3.5 Turbo \cite{openai_gpt35turbo}, a widely used general-purpose model. By comparing these models, we aim to analyze the relative benefits of general-purpose versus code-specialized capabilities, as well as the influence of model scale and architecture on the quality, coherence, and technical fidelity of generated summaries.

\subsubsection{Evaluation Metrics}
To rigorously quantify the quality of generated abstractive summaries, let $Y = \{y^{(i)}\}_{i=1}^N$ denote the set of reference summaries and $\hat{Y} = \{\hat{y}^{(i)}\}_{i=1}^N$ the corresponding set of model-generated summaries for $N$ evaluation instances. Each summary $y^{(i)}$ or $\hat{y}^{(i)}$ is a sequence of tokens in $V_T$, the textual vocabulary.

Our primary metric is \textbf{BERTScore} \cite{zhang2020bertscoreevaluatingtextgeneration}, which measures semantic similarity between generated and reference summaries. Let $E(y)$ denote the sequence of contextual embeddings obtained from a pre-trained transformer for a summary $y$. BERTScore computes a precision, recall, and F1 measure based on pairwise cosine similarity between token embeddings:
\[
\text{Precision}( \hat{y}^{(i)}, y^{(i)}) = \frac{1}{|\hat{y}^{(i)}|} \sum_{ \hat{e} \in E(\hat{y}^{(i)}) } \max_{e \in E(y^{(i)})} \text{cos}(\hat{e}, e),
\]
\[
\text{Recall}( \hat{y}^{(i)}, y^{(i)}) = \frac{1}{|y^{(i)}|} \sum_{ e \in E(y^{(i)}) } \max_{ \hat{e} \in E(\hat{y}^{(i)})} \text{cos}(e, \hat{e}),
\]
\[
\text{F1} = 2 \cdot \frac{\text{Precision} \cdot \text{Recall}}{\text{Precision} + \text{Recall}}.
\]

For comparison with prior work, we also report \textbf{ROUGE-N} and \textbf{ROUGE-L} \cite{lin-2004-rouge}. ROUGE-N measures $n$-gram overlap between the generated and reference summaries:
\[
\text{ROUGE-N} = \frac{\sum_{i=1}^N \sum_{g \in \text{ngrams}_n(y^{(i)})} \min(\text{Count}(g, \hat{y}^{(i)}), \text{Count}(g, y^{(i)}))}{\sum_{i=1}^N \sum_{g \in \text{ngrams}_n(y^{(i)})} \text{Count}(g, y^{(i)})}.
\]
ROUGE-L captures the longest common subsequence (LCS) between the generated and reference sequences:
\[
\text{ROUGE-L} = \frac{\sum_{i=1}^N \text{LCS}(\hat{y}^{(i)}, y^{(i)})}{\sum_{i=1}^N |y^{(i)}|}.
\]

Collectively, these metrics provide a rigorous evaluation of both semantic fidelity (via BERTScore) and surface-level textual alignment (via ROUGE), allowing a comprehensive assessment of abstractive summarization performance.

\begin{figure}
\centering 
\includegraphics[width=0.9\textwidth]{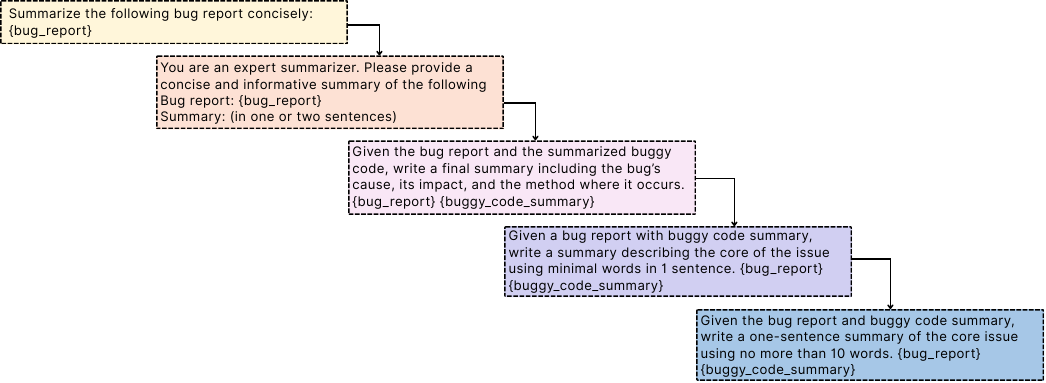} 
 \caption{Overview of the iterative prompt refinement process. Prompts were progressively adjusted based on the quality of model-generated summaries to improve accuracy, completeness, and readability.}
\end{figure}

\subsubsection{Experimental Configurations}
All experiments were performed on a single NVIDIA RTX A6000 GPU within a controlled Docker-based environment to ensure reproducibility. Each input pair $x = (x_T, x_C)$ was tokenized into sequences of tokens from the respective vocabularies $V_T$ and $V_C$, applying truncation or padding as necessary to satisfy the context length limitations of each LLM. We evaluated zero-shot, one-shot, and few-shot prompting strategies, where few-shot prompts incorporated $k$ exemplar input–output pairs concatenated with the target input, formally represented as 
\[
\pi_{\text{fs}}(x) = \langle (\pi(x^{(1)}), y^{(1)}), \dots, (\pi(x^{(k)}), y^{(k)}), \pi(x) \rangle.
\]

For code-containing inputs, sequences exceeding a maximum token length $L_{\max} = 1024$ were partitioned into $M = \lceil |x_C| / L_{\max} \rceil$ contiguous chunks $x_C^{(1)}, \dots, x_C^{(M)}$. Each chunk was summarized individually, producing intermediate summaries $\{s^{(m)}\}_{m=1}^M$ that were then aggregated into a consolidated code-level summary $\tilde{s}$, forming the final input $(x_T, \tilde{s})$ for abstractive summarization.

Text generation from the LLM employed standard decoding strategies including beam search, top-$k$ sampling, and repetition constraints to improve diversity and reduce hallucinations. Fine-tuning was conducted on the Defects4J dataset, using a 90/10 train/validation split. The training configuration included a batch size of 1, gradient accumulation over 4 steps to emulate a larger effective batch size, and a learning rate of $\eta = 5 \times 10^{-5}$. Let $\theta$ denote the model parameters, then the fine-tuning update at step $t$ followed
\[
\theta_{t+1} = \theta_t - \eta \, \nabla_\theta \mathcal{L}(\hat{y}^{(t)}, y^{(t)}),
\]
where $\mathcal{L}$ is the cross-entropy loss between the generated summary $\hat{y}^{(t)}$ and the reference $y^{(t)}$.

\begin{table*}[htb]
\centering
\caption{ROUGE scores of baseline models and LLMs on Fang et al. corpus using bug reports only. Fine-tuned models evaluated on Defects4J; GPT-3.5 Turbo in zero-shot setting.}
\begin{tabular}{l|c|c|c}
\toprule
\textbf{Model} & \textbf{ROUGE-1} & \textbf{ROUGE-2} & \textbf{ROUGE-L} \\
\midrule
SUMLLAMA \cite{xiang2024sumllama} & 0.4426 & 0.2515 & 0.4101 \\
KSCLP \cite{ksclp} & 0.4133 & 0.2202 & 0.3789 \\
RTA \cite{10.1109/ICSE48619.2023.00060} & 0.3919 & 0.2057 & 0.3597 \\
BugSum \cite{bugsum} & 0.2591 & 0.1166 & 0.2469 \\
DeepSum \cite{10.1145/3196321.3196326} & 0.1760 & 0.0805 & 0.1700 \\
\midrule
CodeLlama & 0.2641 & 0.0746 & 0.2206 \\
Llama-3.1 & 0.1220 & 0.0293 & 0.0998 \\
Mistral & \textbf{0.2786} & \textbf{0.0747} & \textbf{0.2316} \\
Phi-3 & 0.1246 & 0.0253 & 0.0988 \\
Gemma & 0.2016 & 0.0425 & 0.1596 \\
Qwen3 & 0.0960 & 0.0189 & 0.0764 \\
DeepSeek Coder & 0.1049 & 0.0202 & 0.0821 \\
GPT-3.5 Turbo & 0.3225 & 0.1146 & 0.2731 \\
\bottomrule
\end{tabular}
\label{tab:baseline_rogue_score}
\end{table*}

\begin{table*}[htb]
\centering
\caption{BERTScore of LLMs on multiple datasets comparing bug reports alone vs bug reports + code; highlights effects of fine-tuning and code integration.}
\resizebox{\textwidth}{!}{
\begin{tabular}{l|l|c|c|c|c|c|c}
\toprule
\textbf{Dataset} & \textbf{Model} & \multicolumn{3}{c|}{\textbf{Bug Reports}} & \multicolumn{3}{c}{\textbf{Bug Reports + Buggy Code}} \\
\cmidrule(lr){3-8}
 &  & \textbf{Precision} & \textbf{Recall} & \textbf{F1 Score} & \textbf{Precision} & \textbf{Recall} & \textbf{F1 Score} \\
\midrule

\multirow{8}{*}{Defects4J \cite{defects4j}}
 & CodeLlama       & 0.6608 & 0.7002 & 0.6767 & 0.5110 & 0.5650 & 0.5308 \\
 & LLaMA-3.1       & 0.4025 & 0.6064 & 0.4816 & 0.3812 & 0.5886 & 0.4588 \\
 & Mistral         & \textbf{0.6764} & \textbf{0.7072} & \textbf{0.6877} & 0.5944 & 0.6220 & 0.6051 \\
 & Phi-3           & 0.4056 & 0.6140 & 0.4849 & 0.4109 & 0.5665 & 0.4728 \\
 & Gemma           & 0.5726 & 0.6756 & 0.6173 & 0.5932 & 0.6463 & 0.6160 \\
 & Qwen3           & 0.4006 & 0.5868 & 0.4744 & 0.3955 & 0.5707 & 0.4658 \\
 & DeepSeek Coder  & 0.4075 & 0.6416 & 0.4958 & 0.4109 & 0.6277 & 0.4933 \\
 & GPT-3.5 Turbo   & 0.7478 & 0.7869 & 0.7640 & - & - & - \\
\midrule

\multirow{8}{*}{SDS \cite{10.1145/1806799.1806872}}
 & CodeLlama       & 0.6072 & 0.3753 & 0.4620 & 0.6049 & 0.3582 & 0.4476 \\
 & LLaMA-3.1       & 0.4969 & 0.4337 & 0.4624 & 0.4947 & 0.4148 & 0.4503 \\
 & Mistral         & \textbf{0.5983} & 0.3539 & 0.4435 & 0.6133 & 0.3662 & \textbf{0.4575} \\
 & Phi-3           & 0.5279 & 0.4001 & 0.4520 & 0.5131 & 0.4150 & 0.4563 \\
 & Gemma           & 0.5821 & 0.3968 & 0.4702 & 0.5871 & 0.3893 & 0.4661 \\
 & Qwen3           & \textbf{0.5532} & \textbf{0.4685} & \textbf{0.5070} & 0.5468 & 0.4673 & 0.5036 \\
 & DeepSeek Coder  & 0.5124 & 0.4394 & 0.4721 & 0.5187 & 0.4442 & 0.4778 \\
 & GPT-3.5 Turbo   & 0.6354 & 0.3745 & 0.4703 & - & - & - \\
\midrule

\multirow{8}{*}{ADS \cite{ADS}}
 & CodeLlama       & 0.6111 & 0.4054 & 0.4856 & 0.6141 & 0.3954 & 0.4795 \\
 & LLaMA-3.1       & 0.4842 & 0.4405 & 0.4599 & 0.4856 & 0.4406 & 0.4607 \\
 & Mistral         & 0.6176 & 0.3977 & 0.4827 & \textbf{0.6294} & 0.4116 & \textbf{0.4967} \\
 & Phi-3           & 0.5165 & 0.4510 & 0.4797 & 0.5075 & 0.4477 & 0.4732 \\
 & Gemma           & 0.5913 & 0.4408 & 0.5041 & 0.5971 & 0.4326 & 0.5004 \\
 & Qwen3           & \textbf{0.5327} & \textbf{0.4884} & \textbf{0.5088} & 0.5351 & \textbf{0.4908} & 0.5113 \\
 & DeepSeek Coder  & 0.5221 & 0.4853 & 0.5016 & 0.5081 & 0.4778 & 0.4910 \\
 & GPT-3.5 Turbo   & 0.6536 & 0.4202 & 0.5104 & - & - & - \\
\midrule

\multirow{8}{*}{Fang's Corpus \cite{10.1109/ICSE48619.2023.00060}}
 & CodeLlama       & 0.6148 & 0.5912 & 0.5991 & 0.6185 & 0.5913 & 0.6010 \\
 & LLaMA-3.1       & 0.3846 & 0.5452 & 0.4476 & 0.3911 & 0.5507 & 0.4542 \\
 & Mistral         & \textbf{0.6297} & \textbf{0.6167} & \textbf{0.6202} & 0.6210 & \textbf{0.6195} & 0.6170 \\
 & Phi-3           & 0.4249 & 0.5645 & 0.4816 & 0.4170 & 0.5673 & 0.4773 \\
 & Gemma           & 0.5581 & 0.6063 & 0.5779 & 0.5746 & 0.6067 & 0.5864 \\
 & Qwen3           & 0.3946 & 0.5484 & 0.4571 & 0.3949 & 0.5485 & 0.4576 \\
 & DeepSeek Coder  & 0.3959 & 0.5801 & 0.4677 & 0.3898 & 0.5749 & 0.4614 \\
 & GPT-3.5 Turbo   & 0.6519 & 0.6330 & 0.6394 & - & - & - \\
\bottomrule
\end{tabular}}
\label{tab:combined_bert_score_ft_results}
\end{table*}

\begin{table*}[htb]
\centering
\caption{BERTScore (Precision, Recall, F1) on Defects4J for bug reports versus bug reports + buggy code across prompting strategies.}
\resizebox{\textwidth}{!}{%
\begin{tabular}{l|l|c|c|c|c|c|c}
\toprule
\textbf{Model} & \textbf{\shortstack{Prompting\\Technique}} & \multicolumn{3}{c|}{\textbf{Bug Reports}} & \multicolumn{3}{c}{\textbf{Bug Reports + Buggy Code}} \\
\cmidrule(lr){3-8}
 &  & \textbf{Precision} & \textbf{Recall} & \textbf{F1 Score} & \textbf{Precision} & \textbf{Recall} & \textbf{F1 Score} \\
\midrule

\multirow{3}{*}{CodeLlama}
 & Zero-Shot & 0.5553 & 0.6408 & 0.5859 & 0.4957 & 0.5350 & 0.5109 \\
 & One-Shot  & 0.4194 & 0.5347 & 0.4618 & 0.5656 & 0.5627 & 0.5615 \\
 & Few-Shot  & 0.6711 & 0.6880 & 0.6752 & 0.6326 & 0.6398 & 0.6337 \\
\midrule

\multirow{3}{*}{Llama-3.1}
 & Zero-Shot & 0.4154 & 0.6015 & 0.4895 & 0.3765 & 0.5742 & 0.4509 \\
 & One-Shot  & 0.3955 & 0.5995 & 0.4745 & 0.3815 & 0.5760 & 0.4561 \\
 & Few-Shot  & 0.4005 & 0.6050 & 0.4795 & 0.5673 & 0.6546 & 0.5974 \\
\midrule

\multirow{3}{*}{Mistral}
 & Zero-Shot & 0.6574 & 0.6924 & 0.6719 & 0.6104 & 0.6212 & 0.6137 \\
 & One-Shot  & 0.6783 & 0.6895 & 0.6816 & 0.6315 & 0.6082 & 0.6180 \\
 & Few-Shot  & 0.6874 & 0.7032 & 0.6928 & 0.6593 & 0.6467 & 0.6509 \\
\midrule

\multirow{3}{*}{Phi-3}
 & Zero-Shot & 0.4311 & 0.6058 & 0.4992 & 0.4965 & 0.5797 & 0.5304 \\
 & One-Shot  & 0.4220 & 0.6299 & 0.5027 & 0.4758 & 0.5209 & 0.4925 \\
 & Few-Shot  & 0.4506 & 0.6176 & 0.5171 & 0.4314 & 0.5767 & 0.4905 \\
\midrule

\multirow{3}{*}{Gemma}
 & Zero-Shot & 0.5662 & 0.6756 & 0.6132 & 0.5271 & 0.6307 & 0.5715 \\
 & One-Shot  & 0.4154 & 0.5862 & 0.4844 & 0.5335 & 0.6054 & 0.5630 \\
 & Few-Shot  & 0.5459 & 0.6463 & 0.5894 & 0.5477 & 0.6161 & 0.5768 \\
\midrule

\multirow{3}{*}{Qwen3}
 & Zero-Shot & 0.4085 & 0.5992 & 0.4846 & 0.4159 & 0.5525 & 0.4709 \\
 & One-Shot  & 0.4135 & 0.6005 & 0.4878 & 0.4202 & 0.5447 & 0.4705 \\
 & Few-Shot  & 0.4059 & 0.5995 & 0.4829 & 0.4102 & 0.5583 & 0.4690 \\
\midrule

\multirow{3}{*}{DeepSeek Coder}
 & Zero-Shot & 0.4132 & 0.6401 & 0.5002 & 0.4080 & 0.5871 & 0.4737 \\
 & One-Shot  & 0.4050 & 0.6308 & 0.4902 & 0.4389 & 0.5880 & 0.4966 \\
 & Few-Shot  & 0.4132 & 0.6401 & 0.5002 & 0.4107 & 0.5987 & 0.4789 \\
\midrule

\multirow{3}{*}{GPT-3.5 Turbo}
 & Zero-Shot & 0.7478 & 0.7869 & 0.7640 & 0.6761 & 0.6763 & 0.6740 \\
 & One-Shot  & 0.8398 & 0.8417 & 0.8388 & 0.7893 & 0.7787 & 0.7820 \\
 & Few-Shot  & \textbf{0.8696} & \textbf{0.8536} & \textbf{0.8595} & \textbf{0.9108} & \textbf{0.8929} & \textbf{0.9003} \\
\bottomrule
\end{tabular}}
\label{tab:models_prompt_bert_score_results}
\end{table*}

\begin{table*}[htb]
\centering
\caption{BERTScore for CodeLlama using only buggy code to assess code-alone contribution.}
\begin{tabular}{l|c|c|c}
\toprule
\textbf{Prompting Technique} & \multicolumn{3}{c}{\textbf{Buggy Code}} \\
\cmidrule(lr){2-4}
 & \textbf{Precision} & \textbf{Recall} & \textbf{F1 Score} \\
\midrule
Zero-Shot & 0.4570 & 0.5005 & 0.4767 \\
\bottomrule
\end{tabular}
\label{tab:bert_score_code_results}
\end{table*}

\begin{table*}[htb]
\centering
\caption{BERTScore for CodeLlama using bug reports + patch code, highlighting the effect of corrected code snippets.}
\begin{tabular}{l|c|c|c}
\toprule
\textbf{Prompting Technique} & \multicolumn{3}{c}{\textbf{Bug Reports + Patch Code}} \\
\cmidrule(lr){2-4}
 & \textbf{Precision} & \textbf{Recall} & \textbf{F1 Score} \\
\midrule
Zero-Shot & 0.5903 & 0.5940 & 0.5888 \\
One-Shot & 0.5613 & 0.5630 & 0.5592 \\
Few-Shot  & \textbf{0.6286} & \textbf{0.6499} & \textbf{0.6368}\\
\bottomrule
\end{tabular}
\label{tab:bert_score_patch_results}
\end{table*}



\begin{table*}[htb]
\centering
\caption{BERTScore comparing input sequences (Code → Bug Report vs Bug Report → Code) using CodeLlama.}
\resizebox{\textwidth}{!}{%
\begin{tabular}{l|c|c|c|c|c|c}
\toprule
\textbf{Prompting Technique} & \multicolumn{3}{c|}{\textbf{Code → Bug Report}} & \multicolumn{3}{c}{\textbf{Bug Report → Code}} \\
\cmidrule(lr){2-7}
 & \textbf{Precision} & \textbf{Recall} & \textbf{F1 Score} & \textbf{Precision} & \textbf{Recall} & \textbf{F1 Score} \\
\midrule
Zero-Shot & \textbf{0.6029} & 0.6087 & 0.6030 & 0.4957 & 0.5350 & 0.5109 \\
One-Shot  & 0.5315 & 0.5667 & 0.5450 & 0.5656 & 0.5627 & 0.5615 \\
Few-Shot  & 0.5992 & \textbf{0.6230} & \textbf{0.6087} & \textbf{0.6326} & \textbf{0.6398} & \textbf{0.6337} \\
\bottomrule
\end{tabular}}
\label{tab:bert_score_input_results}
\end{table*}

\subsection{Baselines}
To evaluate the performance of our proposed approach, we compare against five established bug-report summarization methods. SUMLLAMA \cite{xiang2024sumllama} leverages transformer-based models to generate abstractive summaries using large-scale pretraining. KSCLP \cite{ksclp} applies a knowledge-aware sequence-to-sequence framework, incorporating structural information from both bug reports and software artifacts. RTA \cite{10.1109/ICSE48619.2023.00060} uses a retrieval-augmented summarization approach to enhance coverage of relevant report content. BugSum \cite{bugsum} focuses on extractive summarization with heuristics designed to identify salient sentences in bug reports. DeepSum \cite{10.1145/3196321.3196326} integrates neural sequence modeling with software-specific features for both extractive and abstractive summarization. These baselines provide a diverse set of comparison points, spanning both extractive and abstractive strategies, and enable rigorous evaluation of the advantages offered by multimodal LLM-based summarization.

\section{Experimental Results}
\subsection{Effectiveness Compared to Baselines (RQ1)}
We evaluated the bug-report-only component of our framework against both extractive and abstractive baselines. The extractive baselines include BugSum \cite{bugsum} and DeepSum \cite{10.1145/3196321.3196326}, while the abstractive baselines comprise SUMLLAMA \cite{xiang2024sumllama}, KSCLP \cite{ksclp}, and RTA \cite{10.1109/ICSE48619.2023.00060}. Performance was assessed on the Fang et al. corpus \cite{10.1109/ICSE48619.2023.00060} using ROUGE and BERTScore metrics.

As reported in Table~\ref{tab:baseline_rogue_score}, extractive models underperform on this dataset due to their inability to paraphrase and synthesize novel expressions. BugSum achieved ROUGE-1 = 0.2591, while DeepSum scored 0.1760. In contrast, our models produced more concise and coherent summaries, with Mistral achieving ROUGE-1 = 0.2786, CodeLlama = 0.2641, and Gemma = 0.2016. Although abstractive baselines fine-tuned on the Fang corpus attained higher ROUGE scores—SUMLLAMA (ROUGE-1 = 0.4426, ROUGE-L = 0.4101), KSCLP (ROUGE-1 = 0.4133, ROUGE-L = 0.3789), and RTA (ROUGE-1 = 0.3919, ROUGE-L = 0.3597)—our models maintained strong semantic fidelity, as reflected by BERTScore results in Table~\ref{tab:combined_bert_score_ft_results}.

Quantitatively, the average BERTScore of our models over $N$ evaluation instances consistently exceeded that of extractive baselines demonstrating that even when ROUGE scores are surpassed by fine-tuned abstractive methods, our approach preserves semantic coherence and conciseness.

\begin{tcolorbox}[colback=gray!10,colframe=black,boxrule=0.5pt,arc=2mm,
                  width=\textwidth, boxsep=1pt]
\textbf{Answer to RQ1:} \textit{Our approach generates concise, coherent summaries that outperform extractive baselines and achieve strong semantic performance, remaining competitive with fine-tuned abstractive models.}
\end{tcolorbox}

\subsection{Impact of Code Context (RQ2)}
We evaluated eight LLMs on the Defects4J dataset \cite{defects4j} under two input conditions: using bug reports alone, and using bug reports combined with associated code summaries. The inclusion of code information generally led to modest improvements in few-shot F1 scores. For example, GPT-3.5 Turbo increased from 0.8595 to 0.9003 when code was included, while Mistral showed a decrease from 0.6928 to 0.6509 in certain configurations, as reported in Tables~\ref{tab:models_prompt_bert_score_results}--\ref{tab:bert_score_patch_results}. 

Ablation studies further highlight the influence of code type and input ordering. Using only buggy code yields low F1 performance (0.4767, Table~\ref{tab:bert_score_code_results}), whereas incorporating patch code provides a modest increase (0.6368, Table~\ref{tab:bert_score_patch_results}). The sequence in which inputs are presented to the model also affects results: presenting the bug report first followed by code (Bug Report $\rightarrow$ Code) achieves F1 = 0.6337, while the reverse order (Code $\rightarrow$ Bug Report) decreases performance to 0.6087, as shown in Table~\ref{tab:bert_score_input_results}. These findings indicate that while code can enhance summary quality, bug-report text remains the dominant source of information, and careful consideration of code type and prompt structure is necessary to maximize benefits.

\begin{tcolorbox}[colback=gray!10,colframe=black,boxrule=0.5pt,arc=2mm,
                  width=\textwidth, boxsep=1pt]
\textbf{Answer to RQ2:} \textit{Code integration yields modest gains (F1 up to 0.9003). Bug reports remain primary input; patch code and input order can further influence results. Current metrics may not fully capture qualitative improvements.}
\end{tcolorbox}

\subsection{Effect of Prompt Engineering vs Fine-Tuning (RQ3)}
We analyzed the relative impact of prompt engineering and fine-tuning across multiple datasets. On extractive datasets such as SDS \cite{10.1145/1806799.1806872} and ADS \cite{ADS}, fine-tuning resulted in moderate precision values between 0.5 and 0.6, but recall remained relatively low (0.35--0.45). Incorporating code information in these settings provided minimal additional benefit, indicating that extractive benchmarks are largely driven by textual content in the bug reports.

On abstractive datasets, including Defects4J and Fang et al.'s corpus, fine-tuned models exhibited modest gains. For example, Mistral achieved F1 = 0.6202 without code and 0.6170 with code, while CodeLlama scored 0.5991 without code and 0.6010 with code. Interestingly, zero-shot prompt engineering often matched or outperformed fine-tuned models: GPT-3.5 Turbo achieved F1 = 0.7640 on Defects4J and 0.6394 on the Fang dataset, as summarized in Table~\ref{tab:combined_bert_score_ft_results}. These results suggest that for abstractive summarization, carefully designed prompts can leverage pre-trained knowledge effectively, reducing the need for extensive fine-tuning, particularly when training data is limited.

\begin{tcolorbox}[colback=gray!10,colframe=black,boxrule=0.5pt,arc=2mm,
                  width=\textwidth, boxsep=1pt]
\textbf{Answer to RQ3:} \textit{Fine-tuning yields moderate improvements on extractive datasets but marginal gains on abstractive datasets; prompt engineering remains competitive, particularly for small training datasets.}
\end{tcolorbox}

\section{Limitations \& Future Work}
This study has several limitations. First, the primary dataset, Defects4J, is relatively small, which constrains the diversity of bug-report and code patterns available for model training and evaluation. Second, there is a scarcity of publicly available datasets that provide both bug reports and corresponding code, limiting opportunities for extensive multimodal experimentation. Third, our hierarchical aggregation of long code snippets may result in partial information loss, potentially affecting the fidelity of generated summaries. Fourth, although our prompt engineering strategies were designed to be structured and multi-modal, they are not necessarily optimal, and alternative prompt formulations may yield improved performance.

Future work will address these limitations in several directions. Human evaluation of generated summaries is needed to capture qualitative aspects such as clarity, technical accuracy, and usefulness to developers. Integration of our approach into real-world software maintenance tools could provide practical feedback and validate model utility in situ. Finally, the development of evaluation metrics specifically tailored to bug-report summarization \cite{long2025learningsoftwarebugreports} would enable more precise and domain-relevant assessment of model performance, particularly for capturing semantic fidelity and actionable content beyond standard text-similarity metrics.

\section{Conclusion}
We present a multi-stage LLM-based pipeline for abstractive bug-report summarization that effectively integrates both textual descriptions and associated code snippets. Our approach outperforms extractive baselines, achieving ROUGE improvements ranging from +7.5\% to +58.2\%, while maintaining strong semantic fidelity as measured by BERTScore. Experiments demonstrate that carefully designed prompt engineering can, in some cases, rival or even exceed the benefits of fine-tuning, particularly on limited training data. Despite these promising results, the relatively small size of available datasets and limitations of current evaluation metrics suggest that performance should be interpreted cautiously. Future work targeting human evaluation, larger multimodal datasets, and task-specific metrics will further clarify the practical utility and robustness of LLM-based bug-report summarization.

\bibliographystyle{unsrt}  

\bibliography{references}

@String{Computing = "Computing" }

@String{Computer = "{IEEE} Computer" }

@String{Academic = "Academic Press" }

@ArtifactSoftware{R,
    title = {R: A Language and Environment for Statistical Computing},
    author = {{R Core Team}},
    organization = {R Foundation for Statistical Computing},
    address = {Vienna, Austria},
    year = {2019},
    url = {https://www.R-project.org/},
}

@article{10.1007/s10462-016-9475-9,
author = {Gambhir, Mahak and Gupta, Vishal},
title = {Recent automatic text summarization techniques: a survey},
year = {2017},
issue_date = {January 2017},
publisher = {Kluwer Academic Publishers},
address = {USA},
volume = {47},
number = {1},
issn = {0269-2821},
url = {https://doi.org/10.1007/s10462-016-9475-9},
doi = {10.1007/s10462-016-9475-9},
journal = {Artif. Intell. Rev.},
month = jan,
pages = {1–66},
numpages = {66}
}

@misc{santana2025promptingtechniqueiuse,
      title={Which Prompting Technique Should I Use? An Empirical Investigation of Prompting Techniques for Software Engineering Tasks}, 
      author={E. G. Santana Jr and Gabriel Benjamin and Melissa Araujo and Harrison Santos and David Freitas and Eduardo Almeida and Paulo Anselmo da M. S. Neto and Jiawei Li and Jina Chun and Iftekhar Ahmed},
      year={2025},
      eprint={2506.05614},
      archivePrefix={arXiv},
      primaryClass={cs.SE},
      url={https://arxiv.org/abs/2506.05614}, 
}

@inproceedings{10.1109/ICSE48619.2023.00060,
author = {Fang, Sen and Zhang, Tao and Tan, Youshuai and Jiang, He and Xia, Xin and Sun, Xiaobing},
title = {RepresentThemAll: A Universal Learning Representation of Bug Reports},
year = {2023},
isbn = {9781665457019},
publisher = {IEEE Press},
url = {https://doi.org/10.1109/ICSE48619.2023.00060},
doi = {10.1109/ICSE48619.2023.00060},
booktitle = {Proceedings of the 45th International Conference on Software Engineering},
pages = {602–614},
numpages = {13},
location = {Melbourne, Victoria, Australia},
series = {ICSE '23}
}

@article{xiang2024sumllama,
  author = {Xiang, B. and Shao, Y.},
  title = {SumLLaMA: Efficient Contrastive Representations and Fine-Tuned Adapters for Bug Report Summarization},
  journal = {IEEE Access},
  volume = {12},
  pages = {78562--78571},
  year = {2024},
  doi = {10.1109/ACCESS.2024.3397326}
}

@inproceedings{defects4J-dissection,
    title = {{Dissection of a Bug Dataset: Anatomy of 395 Patches from Defects4J}},
    author = {Sobreira, Victor and Durieux, Thomas and Madeiral, Fernanda and Monperrus, Martin and Maia, Marcelo A.},
    booktitle = {Proceedings of SANER},
    year = {2018},
    doi = {10.1109/SANER.2018.8330203}
}

@inproceedings{10.1145/1453101.1453146,
author = {Bettenburg, Nicolas and Just, Sascha and Schr\"{o}ter, Adrian and Weiss, Cathrin and Premraj, Rahul and Zimmermann, Thomas},
title = {What makes a good bug report?},
year = {2008},
isbn = {9781595939951},
publisher = {Association for Computing Machinery},
address = {New York, NY, USA},
url = {https://doi.org/10.1145/1453101.1453146},
doi = {10.1145/1453101.1453146},
booktitle = {Proceedings of the 16th ACM SIGSOFT International Symposium on Foundations of Software Engineering},
pages = {308–318},
numpages = {11},
location = {Atlanta, Georgia},
series = {SIGSOFT '08/FSE-16}
}

@inproceedings{lin-2004-rouge,
    title = "{ROUGE}: A Package for Automatic Evaluation of Summaries",
    author = "Lin, Chin-Yew",
    booktitle = "Text Summarization Branches Out",
    month = jul,
    year = "2004",
    address = "Barcelona, Spain",
    publisher = "Association for Computational Linguistics",
    url = "https://aclanthology.org/W04-1013/",
    pages = "74--81"
}

@misc{hu2021loralowrankadaptationlarge,
      title={LoRA: Low-Rank Adaptation of Large Language Models}, 
      author={Edward J. Hu and Yelong Shen and Phillip Wallis and Zeyuan Allen-Zhu and Yuanzhi Li and Shean Wang and Lu Wang and Weizhu Chen},
      year={2021},
      eprint={2106.09685},
      archivePrefix={arXiv},
      primaryClass={cs.CL},
      url={https://arxiv.org/abs/2106.09685}, 
}

@misc{zhang2024systematicsurveytextsummarization,
      title={A Systematic Survey of Text Summarization: From Statistical Methods to Large Language Models}, 
      author={Haopeng Zhang and Philip S. Yu and Jiawei Zhang},
      year={2024},
      eprint={2406.11289},
      archivePrefix={arXiv},
      primaryClass={cs.CL},
      url={https://arxiv.org/abs/2406.11289}, 
}

@inproceedings{kang2023large,
  author = {Kang, S. and Yoon, J. and Yoo, S.},
  title = {Large Language Models are Few-shot Testers: Exploring LLM-based General Bug Reproduction},
  booktitle = {Proceedings of IEEE/ACM International Conference on Software Engineering (ICSE)},
  year = {2023},
  arxiv = {2209.11515}
}

@inproceedings{10.1145/1806799.1806872,
author = {Rastkar, Sarah and Murphy, Gail C. and Murray, Gabriel},
title = {Summarizing software artifacts: a case study of bug reports},
year = {2010},
isbn = {9781605587196},
publisher = {Association for Computing Machinery},
address = {New York, NY, USA},
url = {https://doi.org/10.1145/1806799.1806872},
doi = {10.1145/1806799.1806872},
booktitle = {Proceedings of the 32nd ACM/IEEE International Conference on Software Engineering - Volume 1},
pages = {505–514},
numpages = {10},
location = {Cape Town, South Africa},
series = {ICSE '10}
}

@article{summarizingSoftwareartifacts,
  author    = {Najam Nazar and Yan Hu and He Jiang},
  title     = {Summarizing Software Artifacts: A Literature Review},
  journal   = {Journal of Computer Science and Technology},
  volume    = {31},
  number    = {5},
  pages     = {883--909},
  year      = {2016},
  month     = sep,
  doi       = {10.1007/s11390-016-1671-1},
  issn      = {1860-4749},
  url       = {https://doi.org/10.1007/s11390-016-1671-1}
}

@misc{long2025learningsoftwarebugreports,
      title={Learning Software Bug Reports: A Systematic Literature Review}, 
      author={Guoming Long and Jingzhi Gong and Hui Fang and Tao Chen},
      year={2025},
      eprint={2507.04422},
      archivePrefix={arXiv},
      primaryClass={cs.SE},
      url={https://arxiv.org/abs/2507.04422}, 
}

@article{Shakil_2024,
   title={Abstractive text summarization: State of the art, challenges, and improvements},
   volume={603},
   ISSN={0925-2312},
   url={http://dx.doi.org/10.1016/j.neucom.2024.128255},
   DOI={10.1016/j.neucom.2024.128255},
   journal={Neurocomputing},
   publisher={Elsevier BV},
   author={Shakil, Hassan and Farooq, Ahmad and Kalita, Jugal},
   year={2024},
   month=oct, pages={128255} }

@inproceedings{kryscinski-etal-2020-evaluating,
    title = "Evaluating the Factual Consistency of Abstractive Text Summarization",
    author = "Kryscinski, Wojciech  and
      McCann, Bryan  and
      Xiong, Caiming  and
      Socher, Richard",
    editor = "Webber, Bonnie  and
      Cohn, Trevor  and
      He, Yulan  and
      Liu, Yang",
    booktitle = "Proceedings of the 2020 Conference on Empirical Methods in Natural Language Processing (EMNLP)",
    month = nov,
    year = "2020",
    address = "Online",
    publisher = "Association for Computational Linguistics",
    url = "https://aclanthology.org/2020.emnlp-main.750/",
    doi = "10.18653/v1/2020.emnlp-main.750",
    pages = "9332--9346"
}

@inproceedings{10.5555/3495724.3495883,
author = {Brown, Tom B. and Mann, Benjamin and Ryder, Nick and Subbiah, Melanie and Kaplan, Jared and Dhariwal, Prafulla and Neelakantan, Arvind and Shyam, Pranav and Sastry, Girish and Askell, Amanda and Agarwal, Sandhini and Herbert-Voss, Ariel and Krueger, Gretchen and Henighan, Tom and Child, Rewon and Ramesh, Aditya and Ziegler, Daniel M. and Wu, Jeffrey and Winter, Clemens and Hesse, Christopher and Chen, Mark and Sigler, Eric and Litwin, Mateusz and Gray, Scott and Chess, Benjamin and Clark, Jack and Berner, Christopher and McCandlish, Sam and Radford, Alec and Sutskever, Ilya and Amodei, Dario},
title = {Language models are few-shot learners},
year = {2020},
isbn = {9781713829546},
publisher = {Curran Associates Inc.},
address = {Red Hook, NY, USA},
booktitle = {Proceedings of the 34th International Conference on Neural Information Processing Systems},
articleno = {159},
numpages = {25},
location = {Vancouver, BC, Canada},
series = {NIPS '20}
}

@article{jiang-etal-2020-know,
    title = "How Can We Know What Language Models Know?",
    author = "Jiang, Zhengbao  and
      Xu, Frank F.  and
      Araki, Jun  and
      Neubig, Graham",
    editor = "Johnson, Mark  and
      Roark, Brian  and
      Nenkova, Ani",
    journal = "Transactions of the Association for Computational Linguistics",
    volume = "8",
    year = "2020",
    address = "Cambridge, MA",
    publisher = "MIT Press",
    url = "https://aclanthology.org/2020.tacl-1.28/",
    doi = "10.1162/tacl_a_00324",
    pages = "423--438"
}

@misc{zhang2020bertscoreevaluatingtextgeneration,
      title={BERTScore: Evaluating Text Generation with BERT}, 
      author={Tianyi Zhang and Varsha Kishore and Felix Wu and Kilian Q. Weinberger and Yoav Artzi},
      year={2020},
      eprint={1904.09675},
      archivePrefix={arXiv},
      primaryClass={cs.CL},
      url={https://arxiv.org/abs/1904.09675}, 
}

@inproceedings{defects4j,
  author = {Just, R. and Jalali, D. and Ernst, M. D.},
  title = {Defects4J: A Database of Existing Faults to Enable Controlled Testing Studies for Java Programs},
  booktitle = {Proceedings of the 2014 International Symposium on Software Testing and Analysis (ISSTA 2014)},
  year = {2014},
  pages = {437--440},
  publisher = {ACM},
  doi = {10.1145/2610384.2628055},
  url = {https://doi.org/10.1145/2610384.2628055}
}

@article{ADS,
  author    = {He Jiang and Jingxuan Zhang and Hongjing Ma and Najam Nazar and Zhilei Ren},
  title     = {Mining authorship characteristics in bug repositories},
  journal   = {Science China Information Sciences},
  volume    = {60},
  number    = {1},
  pages     = {012107},
  year      = {2017},
  doi       = {10.1007/s11432-014-0372-y},
  url       = {https://link.springer.com/article/10.1007/s11432-014-0372-y},
  note      = {Accessed: 2025-08-31}
}

@article{SDS,
  author    = {He Jiang and Xiaochen Li and Zhilei Ren and Jifeng Xuan and Zhi Jin},
  title     = {Towards Better Summarizing Bug Reports with Crowdsourcing Elicited Attributes},
  journal   = {arXiv preprint arXiv:1810.00125},
  year      = {2018},
  url       = {https://arxiv.org/abs/1810.00125},
  note      = {Accessed: 2025-08-31}
}

@article{10.1145/3649828,
author = {Li, Haonan and Hao, Yu and Zhai, Yizhuo and Qian, Zhiyun},
title = {Enhancing Static Analysis for Practical Bug Detection: An LLM-Integrated Approach},
year = {2024},
issue_date = {April 2024},
publisher = {Association for Computing Machinery},
address = {New York, NY, USA},
volume = {8},
number = {OOPSLA1},
url = {https://doi.org/10.1145/3649828},
doi = {10.1145/3649828},
journal = {Proc. ACM Program. Lang.},
month = apr,
articleno = {111},
numpages = {26}
}

@INPROCEEDINGS{10764843,
  author={Bo, Lili and Ji, Wangjie and Sun, Xiaobing and Zhang, Ting and Wu, Xiaoxue and Wei, Ying},
  booktitle={2024 39th IEEE/ACM International Conference on Automated Software Engineering (ASE)}, 
  title={ChatBR: Automated Assessment and Improvement of Bug Report Quality Using ChatGPT}, 
  year={2024},
  volume={},
  number={},
  pages={1472-1483},
  doi={}}

@INPROCEEDINGS{7884638,
  author={Chatterjee, Preetha and Nishi, Manziba Akanda and Damevski, Kostadin and Augustine, Vinay and Pollock, Lori and Kraft, Nicholas A.},
  booktitle={2017 IEEE 24th International Conference on Software Analysis, Evolution and Reengineering (SANER)}, 
  title={What information about code snippets is available in different software-related documents? An exploratory study}, 
  year={2017},
  volume={},
  number={},
  pages={382-386},
  doi={10.1109/SANER.2017.7884638}}

@inproceedings{FSE086,
  author    = {Senthil Mani and Anush Sankaran and Rahul Aralikatte},
  title     = {DeepTriage: Exploring the Effectiveness of Deep Learning for Bug Triaging},
  booktitle = {Proceedings of the 40th ACM/IEEE International Conference on Software Engineering (ICSE)},
  year      = {2018},
  pages     = {602--614},
  publisher = {IEEE},
  doi       = {10.1109/ICSE.2018.00076},
  url       = {https://doi.org/10.1109/ICSE.2018.00076},
  note      = {Accessed: 2025-08-31}
}

@article{ksclp,
  author    = {Shao, Y. and Zhang, T. and Tan, Y. and Jiang, H. and Xia, X. and Sun, X.},
  title     = {Towards Effective Bug Report Summarization by Domain-Specific Representation Learning},
  journal   = {arXiv preprint arXiv:2409.00630},
  year      = {2024},
  url       = {https://arxiv.org/abs/2409.00630},
  note      = {Accessed: 2025-08-31}
}

@inproceedings{bugsum,
  author    = {Li, X. and Zhang, J. and Ma, H. and Nazar, N. and Ren, Z.},
  title     = {Deep Context Understanding for Bug Report Summarization},
  booktitle = {Proceedings of the 22nd International Conference on Program Comprehension (ICPC)},
  year      = {2018},
  pages     = {1--10},
  publisher = {IEEE},
  doi       = {10.1145/3196321.3196334},
  url       = {https://dl.acm.org/doi/10.1145/3196321.3196334},
  note      = {Accessed: 2025-08-31}
}

@article{ELKASSAS2021113679,
title = {Automatic text summarization: A comprehensive survey},
journal = {Expert Systems with Applications},
volume = {165},
pages = {113679},
year = {2021},
issn = {0957-4174},
doi = {https://doi.org/10.1016/j.eswa.2020.113679},
url = {https://www.sciencedirect.com/science/article/pii/S0957417420305030},
author = {Wafaa S. El-Kassas and Cherif R. Salama and Ahmed A. Rafea and Hoda K. Mohamed}
}

@ARTICLE{8466000,
  author={Zou, Weiqin and Lo, David and Chen, Zhenyu and Xia, Xin and Feng, Yang and Xu, Baowen},
  journal={IEEE Transactions on Software Engineering}, 
  title={How Practitioners Perceive Automated Bug Report Management Techniques}, 
  year={2020},
  volume={46},
  number={8},
  pages={836-862},
  doi={10.1109/TSE.2018.2870414}}

@inproceedings{10.1145/1718918.1718973,
author = {Breu, Silvia and Premraj, Rahul and Sillito, Jonathan and Zimmermann, Thomas},
title = {Information needs in bug reports: improving cooperation between developers and users},
year = {2010},
isbn = {9781605587950},
publisher = {Association for Computing Machinery},
address = {New York, NY, USA},
url = {https://doi.org/10.1145/1718918.1718973},
doi = {10.1145/1718918.1718973},
booktitle = {Proceedings of the 2010 ACM Conference on Computer Supported Cooperative Work},
pages = {301–310},
numpages = {10},
location = {Savannah, Georgia, USA},
series = {CSCW '10}
}

@inproceedings{10.1145/3196321.3196326,
author = {Li, Xiaochen and Jiang, He and Liu, Dong and Ren, Zhilei and Li, Ge},
title = {Unsupervised deep bug report summarization},
year = {2018},
isbn = {9781450357142},
publisher = {Association for Computing Machinery},
address = {New York, NY, USA},
url = {https://doi.org/10.1145/3196321.3196326},
doi = {10.1145/3196321.3196326},
booktitle = {Proceedings of the 26th Conference on Program Comprehension},
pages = {144–155},
numpages = {12},
location = {Gothenburg, Sweden},
series = {ICPC '18}
}

@article{team2024gemma,
  title={Gemma: Open models based on gemini research and technology},
  author={Team, Gemma and Mesnard, Thomas and Hardin, Cassidy and Dadashi, Robert and Bhupatiraju, Surya and Pathak, Shreya and Sifre, Laurent and Rivi{\`e}re, Morgane and Kale, Mihir Sanjay and Love, Juliette and others},
  journal={arXiv preprint arXiv:2403.08295},
  year={2024}
}

@misc{rozière2024codellamaopenfoundation,
      title={Code Llama: Open Foundation Models for Code}, 
      author={Baptiste Rozière and Jonas Gehring and Fabian Gloeckle and Sten Sootla and Itai Gat and Xiaoqing Ellen Tan and Yossi Adi and Jingyu Liu and Romain Sauvestre and Tal Remez and Jérémy Rapin and Artyom Kozhevnikov and Ivan Evtimov and Joanna Bitton and Manish Bhatt and Cristian Canton Ferrer and Aaron Grattafiori and Wenhan Xiong and Alexandre Défossez and Jade Copet and Faisal Azhar and Hugo Touvron and Louis Martin and Nicolas Usunier and Thomas Scialom and Gabriel Synnaeve},
      year={2024},
      eprint={2308.12950},
      archivePrefix={arXiv},
      primaryClass={cs.CL},
      url={https://arxiv.org/abs/2308.12950}, 
}

@article{grattafiori2024llama,
  title={The llama 3 herd of models},
  author={Grattafiori, Aaron and Dubey, Abhimanyu and Jauhri, Abhinav and Pandey, Abhinav and Kadian, Abhishek and Al-Dahle, Ahmad and Letman, Aiesha and Mathur, Akhil and Schelten, Alan and Vaughan, Alex and others},
  journal={arXiv preprint arXiv:2407.21783},
  year={2024}
}

@misc{jiang2023mistral7b,
      title={Mistral 7B}, 
      author={Albert Q. Jiang and Alexandre Sablayrolles and Arthur Mensch and Chris Bamford and Devendra Singh Chaplot and Diego de las Casas and Florian Bressand and Gianna Lengyel and Guillaume Lample and Lucile Saulnier and Lélio Renard Lavaud and Marie-Anne Lachaux and Pierre Stock and Teven Le Scao and Thibaut Lavril and Thomas Wang and Timothée Lacroix and William El Sayed},
      year={2023},
      eprint={2310.06825},
      archivePrefix={arXiv},
      primaryClass={cs.CL},
      url={https://arxiv.org/abs/2310.06825}, 
}

@article{abdin2024phi,
  title={Phi-4 technical report},
  author={Abdin, Marah and Aneja, Jyoti and Behl, Harkirat and Bubeck, S{\'e}bastien and Eldan, Ronen and Gunasekar, Suriya and Harrison, Michael and Hewett, Russell J and Javaheripi, Mojan and Kauffmann, Piero and others},
  journal={arXiv preprint arXiv:2412.08905},
  year={2024}
}

@article{achiam2023gpt,
  title={Gpt-4 technical report},
  author={Achiam, Josh and Adler, Steven and Agarwal, Sandhini and Ahmad, Lama and Akkaya, Ilge and Aleman, Florencia Leoni and Almeida, Diogo and Altenschmidt, Janko and Altman, Sam and Anadkat, Shyamal and others},
  journal={arXiv preprint arXiv:2303.08774},
  year={2023}
}

@article{yang2025qwen3,
  title={Qwen3 technical report},
  author={Yang, An and Li, Anfeng and Yang, Baosong and Zhang, Beichen and Hui, Binyuan and Zheng, Bo and Yu, Bowen and Gao, Chang and Huang, Chengen and Lv, Chenxu and others},
  journal={arXiv preprint arXiv:2505.09388},
  year={2025}
}

@misc{openai_gpt35turbo,
  author       = {OpenAI},
  title        = {GPT-3.5 Turbo},
  year         = {2023},
  howpublished = {\url{https://platform.openai.com/docs/models/gpt-3.5-turbo}},
  note         = {Accessed: 2025-09-11}
}

@misc{guo2024deepseekcoderlargelanguagemodel,
      title={DeepSeek-Coder: When the Large Language Model Meets Programming -- The Rise of Code Intelligence}, 
      author={Daya Guo and Qihao Zhu and Dejian Yang and Zhenda Xie and Kai Dong and Wentao Zhang and Guanting Chen and Xiao Bi and Y. Wu and Y. K. Li and Fuli Luo and Yingfei Xiong and Wenfeng Liang},
      year={2024},
      eprint={2401.14196},
      archivePrefix={arXiv},
      primaryClass={cs.SE},
      url={https://arxiv.org/abs/2401.14196}, 
}

@inproceedings{10.5555/3698900.3698946,
author = {Liu, Peiyu and Liu, Junming and Fu, Lirong and Lu, Kangjie and Xia, Yifan and Zhang, Xuhong and Chen, Wenzhi and Weng, Haiqin and Ji, Shouling and Wang, Wenhai},
title = {Exploring ChatGPT's capabilities on vulnerability management},
year = {2024},
isbn = {978-1-939133-44-1},
publisher = {USENIX Association},
address = {USA},
booktitle = {Proceedings of the 33rd USENIX Conference on Security Symposium},
articleno = {46},
numpages = {18},
location = {Philadelphia, PA, USA},
series = {SEC '24}
}
\end{document}